\documentclass[twocolumn,showpacs,superscriptaddress,floatfix]{revtex4-1}
\usepackage{amsmath}
\usepackage{graphicx}
\usepackage{color}
\usepackage{amstext}
\usepackage{hyperref}

 \renewcommand{\Re}{\mathrm{\text{Re}}}
\newcommand{\eps}{\varepsilon}

\begin{document}

\title{Dynamical landscape of transitional pipe flow}

\author{Anna \surname{Frishman}}
\affiliation{Technion Israel Institute of Technology, 32000 Haifa,
  Israel}
\author{Tobias \surname{Grafke}}
\affiliation{University of Warwick, Coventry CV4 7AL, United Kingdom}

\date{\today}

\begin{abstract}
  The transition to turbulence in pipes is characterized by a
  coexistence of laminar and turbulent states. At the lower end of the
  transition, localized turbulent pulses, called puffs, can be
  excited. Puffs can decay when rare fluctuations drive them close to
  an edge state lying at the phase-space boundary with laminar
  flow. At higher Reynolds numbers, homogeneous turbulence can be
  sustained, and dominates over laminar flow. Here we complete this
  landscape of localized states, placing it within a unified
  bifurcation picture. We demonstrate our claims within the Barkley
  model, and motivate them generally. Specifically, we suggest the
  existence of an antipuff and a gap-edge---states which mirror the
  puff and related edge state. Previously observed laminar gaps
  forming within homogeneous turbulence are then naturally identified
  as antipuffs nucleating and decaying through the gap edge.
\end{abstract}



\maketitle

In pipe flow, turbulence first appears intermittently in space,
interspersed with laminar flow, rather than homogeneously in the
entire pipe~\cite{reynolds:1883,lindgren:1957,wygnanski:1973}. This is
characteristic of the subcritical transition to turbulence in wall
bounded flows where turbulence coexists with the linearly stable
laminar flow (Hagen-Poiseuille profile for
pipes)~\cite{manneville:2016}. Thus, turbulence can be excited only
through a large enough perturbation of the base flow. At the low end
of the transitional regime, controlled by the Reynolds number
$\text{Re}$, such excitations generically develop into a localized
turbulent patch, called a puff for pipe flow. Initially, puffs have
short lifetimes and tend to rapidly decay. As $\text{Re}$ increases,
puffs become increasingly stable to decays, but puff splitting, a
single puff turning into two, becomes increasingly more likely,
allowing the proliferation of turbulence~\cite{avila:2011}. Then, at
high enough $\text{Re}$ (termed $\text{Re}_{\text{slug}}$ here) puffs
are replaced by expanding turbulent structures, called slugs, with
laminar flashes randomly opening and closing within their turbulent
cores. This is the regime of intermittent
turbulence~\cite{moxey:2010}: a homogeneous state where turbulence
production matches turbulence dissipation can occupy the entire pipe,
but coexists with random laminar pockets. Further increasing the
Reynolds number, such flashes make way to a homogeneous turbulent core
within the slug, ending the transitional regime.

There are three key states around which the coarse grained dynamics
are known to be organized below $\text{Re}_{\text{slug}}$: the laminar
base flow, the (chaotic) puff state and a state called the edge state,
here termed the \textit{decay edge}, which controls puff excitations
and decays. Even above $\text{Re}_{\text{slug}}$, it is known that the
decay edge remains surprisingly
unchanged~\cite{mellibovsky:2009,duguet:2010}. In this paper we expand
this phase space of states, proposing novel states together with
their bifurcations with $\text{Re}$. These novel states, the
\textit{gap edge} and \textit{antipuff}, mirror the decay edge and
puff, playing an analogous role for the intermittent turbulence above
$\text{Re}_{\text{slug}}$. In addition, the suggested bifurcation
diagram clarifies how the puff state can disappear while the decay
edge remains. Thus, a unified picture of the transitional regime
emerges, demonstrating how this regime can be fruitfully interpreted
in a dynamical systems framework.  We argue for the proposed picture
on general grounds and verify its validity using the Barkley
model~\cite{barkley:2016}.

\section{Background}

Here we provide further details about the puff and decay edge and the
corresponding phase space structure. We also introduce the coarse
grained dynamical point of view taken in the
following~\cite{barkley:2016}, and motivate our use of the Barkley
model.

A puff is a localized chaotic traveling wave, which, while having a
long lifetime, is only of transient nature, forming a chaotic saddle
in phase space~\cite{eckhardt:2007}. Considering localized structures,
phase space can be roughly separated into initial conditions which
directly laminarize, and those which decay after a long transient,
visiting the puff state
first~\cite{skufca:2006,schneider:2007}. Separating these two sets is
the so called \textit{edge of chaos}, small perturbations around which
end up either in the laminar or the puff state. Furthermore, the edge
of chaos corresponds to the stable manifold of the decay edge
state~\cite{skufca:2006,mellibovsky:2009}, an attracting state for
trajectories on the edge which has a single transverse unstable
direction. It leads to a puff state on the one side of the edge and to
the laminar state on the other. The decay edge and the puff share a
similar spatial structure, and there is evidence that they originate
in a saddle node bifurcation at a lower
$\text{Re}$~\cite{mellibovsky:2009,avila:2013}.

The point of view taken here is to treat the puff, decay edge and
homogeneous turbulence as well defined dynamical states, characterized
by an average structure. This is a coarse grained view~\cite{pomeau:2015}, wherein the
detailed chaotic dynamics are treated as noise around the average
state. Thus, while the chaotic dynamics themselves have a rich
dynamical structure, organized around unstable solutions of the
governing equations~\cite{faisst:2003,kerswell:2005,gibson:2008,kawahara:2012}, as
evidenced both for the puff and the decay
edge~\cite{duguet:2008,duguet:2008a,willis:2013,avila:2013,budanur:2017},
we focus on a coarser dynamical description.

Following~\cite{barkley:2015,barkley:2016,song:2017} we focus on two
variables meant to capture the state of the flow at a cross section of
the pipe, and which can vary along the pipe direction $x$. Namely, the
mean shear $u(x,t)$ and turbulent velocity fluctuations
$q(x,t)$. Turbulent fluctuations could be captured through the
transverse velocity root-mean-square, averaged over the pipe cross
section~\cite{song:2017}, being zero in the laminar state. A proxy for
the mean shear is the local centerline velocity: it is smallest in a
turbulent flow where the mean profile is almost flat---equal to the
mean flow rate $u\approx \bar{U}$ ($\bar{U}$ is also called the bulk
velocity), and largest for the base laminar Hagen-Poiseuille flow,
with $u=U_0=2\bar{U}$.  The mean flow shear and the turbulence level
are the minimum ingredients required to capture the dynamical
processes behind turbulence generation and its
sustainment~\cite{pope:2000}. Moreover, based on these two variables
the Barkley model successfully reproduces both qualitative and
quantitative features of pipe, as well as duct
flow~\cite{barkley:2015}. The stochastic version of the Barkley model
further displays the phenomenology of puff splitting and decay in pipe
flows, as well as the intermittent turbulence
regime~\cite{barkley:2016}.

\begin{figure}[b]
  \begin{center}
    \includegraphics[width=0.9\columnwidth]{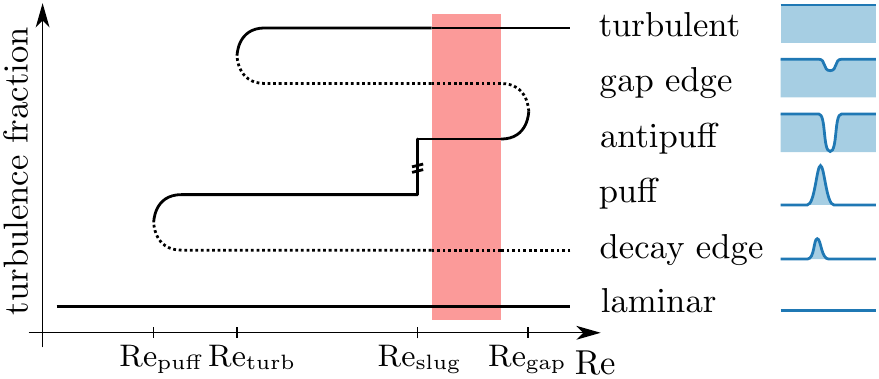}
  \end{center}
  \caption{Sketch of the bifurcation diagram for transitional pipe
    flow. Attracting states are solid lines, unstable edge states are
    dotted. In the deterministic Barkley model used here
    $r_{\text{turb}}=0.667,r_{\text{slug}}=0.726,r_{\text{gap}}=0.736$
  } \label{fig:phase-diagram}
\end{figure}

The key insight at the heart of the Barkley model is that the
transition from puffs to slugs is a transition from an excitable
system to a bi-stable system: turbulence can be excited but not
sustained below $\text{Re}_{\text{slug}}$, whereas homogeneous
turbulence, with spatially uniform turbulence level and mean shear
$(q_t,u_t)$, coexists with laminar flow $(0,U_0)$ as a stable state
above $\text{Re}_{\text{slug}}$. An important feature, which the model
reproduces, is a continuous transition from slugs to
puffs~\cite{duguet:2010}, interpreted as a "masked transition": the
homogeneous turbulent state actually first appears at a $\text{Re}$
below $\text{Re}_{\text{slug}}$, denoted here by
$\text{Re}_{\text{turb}}$, but is masked by the presence of
puffs~\cite{barkley:2015}. This completes the known part of the
bifurcation diagram for the transitional regime which we expand in the
following, see Fig.~\ref{fig:phase-diagram}.

\section{A unified bifurcation diagram}

We propose two novel states which complete the set of basic states in the
transitional region, Fig.~\ref{fig:phase-diagram}: the \emph{gap edge}
and \emph{antipuff}. These are traveling wave states, consisting of
localized laminar flow embedded within homogeneous turbulence. In the
region $\text{Re}_{\text{turb}}<\text{Re}<\text{Re}_{\text{slug}}$ the
gap edge is an unstable state lying at the edge between sustained
homogeneous turbulence and localized turbulence in the form of a puff,
analogously to the decay edge separating the base laminar flow and the
puff. Above $\text{Re}_{\text{slug}}$ puffs disappear but the gap edge
remains, separating homogeneous turbulence from a stable laminar
pocket state we call an antipuff, which is the mirror image of a
puff. Note that at $\text{Re}_{\text{slug}}$, slugs neither expand
nor contract, corresponding to multiple solutions with sections of
arbitrary length at the turbulent and laminar fixed points, which can
be interpreted either as puffs or antipuffs, represented by a
vertical line in Fig.~\ref{fig:phase-diagram}. Finally, the gap edge
and the antipuff disappear together at $\text{Re}_{\text{gap}}$. We
propose that the intermittent turbulence regime observed in pipe flow
corresponds to the random excitations and decays of antipuffs through
the decay edge, and thus $\text{Re}_{\text{gap}}$ marks the end of
this regime. A connection of the observed laminar pockets, here
interpreted as antipuffs, to the laminar tails of slugs has been
previously recognized~\cite{moxey:2010,barkley:2016}, though their
existence as distinct stable structures not explicitly stated. We now
substantiate this picture and flesh out the conditions for its
validity.

\subsection{General considerations}

A key characteristic of puffs are fronts: spatial locations where,
while $u$ remains roughly constant, the turbulence level, $q$, either
sharply rises from zero to a finite value (the upstream front with
$u=U_0$) or sharply decreases to zero from a finite value (downstream,
with $u<U_0$). The front speeds determine the speed of puffs and the
$\text{Re}$ range for their existence. Analogously, front speeds play
a key role in establishing the existence of \emph{antipuffs}.  We
denote by $c_{+}(u,\text{Re})$ ($c_{-}(u,\text{Re})$) the front speed
at mean velocity $u$ where the turbulence level increases (decreases)
in the downstream direction. Turbulence has been shown to be advected
with speed $u-\zeta$ in pipe flow~\cite{song:2017}, where $\zeta$ is a
constant offset velocity from the centerline value. Writing
$c_{-}(u,\text{Re})=u-\zeta+S(u,\text{Re})$, the relative speed
$S(u,\text{Re})$ thus determines the relative stability of laminar
flow ($q=0$) compared with a turbulent flow ($q\neq 0$) at a common
velocity $u$.  Indeed, if $S(u,\text{Re})<0$ the downstream laminar
flow overtakes the upstream turbulent flow, which is thus less stable
at this $u$~\cite{pomeau:1986}. As $c_{+}(u,\text{Re})$ represents the
same physics but with turbulence downstream of laminar flow,
$c_{+}(u,\text{Re})=u-\zeta-S(u,\text{Re})$~\footnote{This is only
  roughly true: the mean velocity profile advecting the turbulence
  creates an a-symmetry between the two fronts, which a 1D model
  cannot capture}. Puffs exist as long as front speeds match: there
exists $u_p$ such that $c_{-}(u_p,\text{Re})=c_{+}(U_0,\text{Re})$. At
$\text{Re}>\text{Re}_{\text{slug}}$, $u_p<u_t$, where $u_t$ is the
homogeneous turbulence mean flow. Puffs are replaced by weak slugs,
which have a downstream front at the turbulent velocity $u_t$.  Since
$c_{-}(u_t,\text{Re})>c_{+}(U_0,\text{Re})$ slugs expand. Generally,
$S(u,\text{Re})$ is an increasing function of $u$ and $\text{Re}$: the
higher the shear, the higher the production of turbulence; the higher
the $\text{Re}$ the lower the dissipation of turbulence---both making
turbulence more sustainable.

The condition for existence of antipuffs is a region where
$S(u_t,\text{Re})<0$ for $\text{Re}>\text{Re}_{\text{slug}}$,
satisfied in pipe flows for $\text{Re}\in
(2250,3000)$~\cite{song:2017}. Indeed, starting from a fully turbulent
pipe flow, $q=q_t, u=u_t$, imagine a local decrease of the level of
turbulence to zero in a small interval in the pipe, while keeping
$u=u_t$. This forms two fronts back to back, with relative speed
$c_+(u_t,\text{Re})-c_-(u_t,\text{Re})=-2S(u_t,\text{Re})>0$ producing
an initially expanding laminar region. The flat turbulent profile,
however, cannot be sustained at $q=0$, and $u$ will relax towards
$U_0$. If $u$ were to reach $U_0$, forming an upstream front of a
slug, then the gap would tend to close since
$c_{+}(u_t,\text{Re})-c_{-}(U_0,\text{Re})<0$. Thus, there exists a
velocity $u_t<u_{\text{ap}}<U_0$, the antipuff speed, giving matching
front speeds $c_-(u_t,\text{Re})=c_+(u_{\text{ap}},\text{Re})$ which
define the antipuff.  At $\text{Re}=\text{Re}_{\text{slug}}$ puff
fronts satisfy
$c_-(u_t,\text{Re}_{\text{slug}})=c_+(U_0,\text{Re}_{\text{slug}})$,
so that $u_{\text{ap}}=U_0$ is a solution for antipuff
fronts. Assuming it is unique, then
$\text{Re}<\text{Re}_{\text{slug}}$ gives $u_{\text{ap}}>U_0$ and
antipuffs disappear. At the other end, antipuffs merge with the
\emph{gap edge} and disappear once $S(u_t,\text{Re})=0$, occurring at
$\text{Re}=\text{Re}_{\text{gap}}$. To motivate the existence and
structure of the gap edge, consider reducing $q$ locally in a
turbulent pipe keeping $u=u_t$: the level of turbulence will return to
$q_t$ if reduced by a minuscule amount, homogeneous turbulence being
stable, while setting $q=0$ will open a laminar pocket which will
expand into an antipuff (or puff, depending on $\text{Re}$).  Thus,
there exists an intermediate value of turbulence $0<q_g<q_t$ right at
the boundary, allowing for a traveling wave solution with upstream
$q_t\to q_g$, and downstream $q_g\to q_t$ fronts at almost the same
speed $u\approx u_t$ (due to slow adjustment of $u$ to $q$), traveling
at speed close to $u_t-\zeta$.

\section{The Barkley model}

We now turn to the Barkley model, describing the numerical results we
have obtained in support of the above described picture, as well as
some asymptotic analytical results.  The dynamics in the Barkley model
reads
\begin{equation}
  \label{eq:model}
  \begin{cases}
    \partial_t q +(u-\zeta)\partial_x q=
    f_r(q,u)+D\partial_x^2q+\sigma q \eta \\ \partial_t u +u\partial_x
    u= \epsilon\left[ (U_0-u)+\kappa(\bar{U}-u)q\right]\\
  \end{cases}
\end{equation}
with $f_r(q,u)=q(r+u-U_0-(r+\delta)(q-1)^2)$.  Velocities are
normalized such that $\bar{U}=1$ and $U_0=2$. The parameter $r$ plays
the role of $\text{Re}$ and $\eta$ is a spatiotemporal white noise
with strength $\sigma$, modeling chaotic fluctuations. While the real
turbulent states are chaotic and spatially intricate, the essential
dynamical and physical features in the transitional region are very
well captured within the Barkley
model~\cite{barkley:2015,barkley:2016}.

\subsection{Results for the Barkley model}

\begin{figure}
  \begin{center}
    \includegraphics[width=\columnwidth]{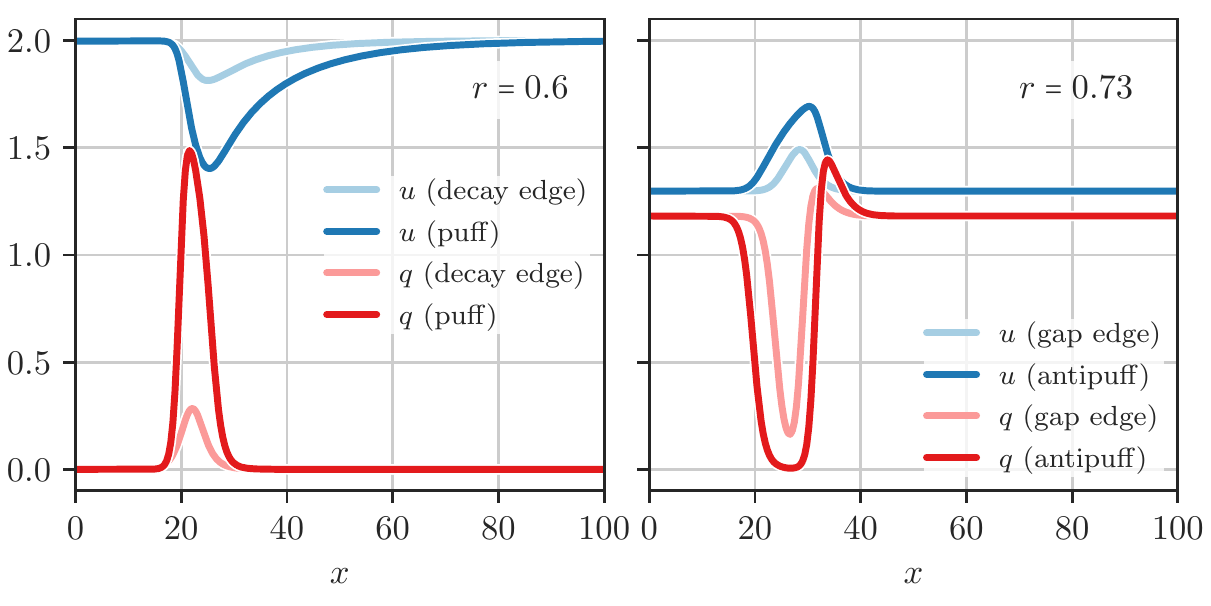}
  \end{center}
  \caption{Spatial profiles of turbulence level $q$ and mean velocity
    $u$ in the Barkley model: puff and decay edge at
    $r<r_{\text{turb}}$ (left), and antipuff and gap edge at
    $r_{\text{slug}}<r<r_{\text{gap}}$
    (right).} \label{fig:deterministic-states}
\end{figure}

We first present the new states, obtained numerically for the model
and then provide the details for the numerical methodology. The
spatial profile of an antipuff as well as that of the gap edge, the
latter obtained by edge tracking, are shown in
Fig.~\ref{fig:deterministic-states} for a representative value of
$r$. Note that while the turbulence drops to zero inside the
antipuff, the centerline velocity $u$ does not reach the laminar
value of $2$, consistent with observations in pipe
flow~\cite{moxey:2010}.

\begin{figure}
  \begin{center}
    \includegraphics[width=\linewidth]{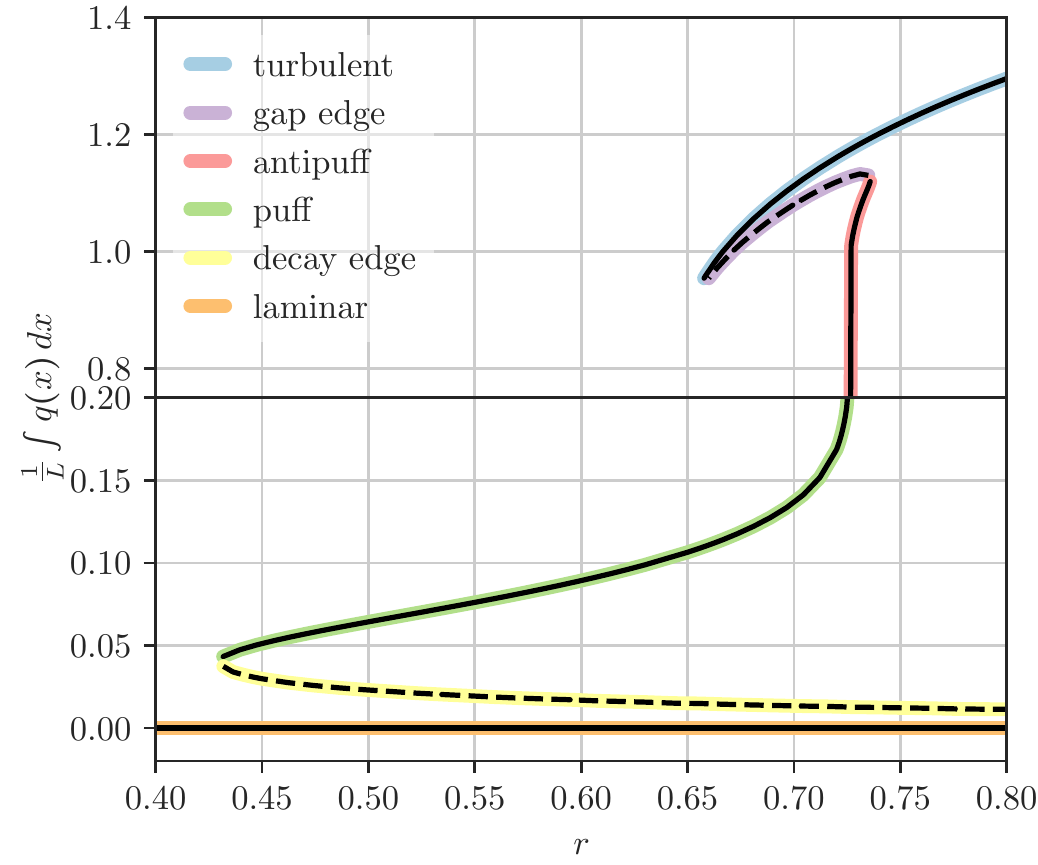}
  \end{center}
  \caption{Explicitly computed bifurcation diagram for the Barkley
    model. The stable states for each $r$ (laminar, turbulent, puff,
    antipuff) are computed by relaxation of the dynamics. The
    unstable edge states (gap edge, decay edge) are computed by the
    edge tracking algorithm described in
    section~\ref{sec:edge-track-algor}. Note the jump in the $y$-axis
    from 0.2 to 0.8 at the solid horizontal grid
    line.\label{fig:phase-diagram-model}}
\end{figure}

The full bifurcation diagram is shown in
Fig.~\ref{fig:phase-diagram-model}, where states are ordered by their
turbulent mass. The measured bifurcations for the Barkley model are
exactly those sketched in Fig.~\ref{fig:phase-diagram}. Note the gap
in turbulent mass formed between the turbulent state and the gap edge
with increasing $r$, and the eventual merging of the gap edge and
antipuff as expected.

\subsection{Methodology for numerical experiments}

For the numerical experiments using the Barkley
model~(\ref{eq:model}), we use the same parameters as in
Ref.~\cite{barkley:2016}: $\zeta=0.8$, $\delta=0.1$, $\eps_1 = 0.1$,
$\eps_2=0.2$, $U_0=2$, $\bar U=1$, $D=0.5$, and $x\in[0,L]$ periodic
with $L=100$. Space is discretized with $N_x=128$ or $N_x=256$ grid
points, and spatial derivatives are computed via fast Fourier
transforms. Temporal integration is performed by a first-order
exponential time differencing (ETD) scheme~\cite{cox-matthews:2002},
with time-steps between $\Delta t=10^{-2}$ and $\Delta t=10^{-3}$. In
simulations including stochastic noise, we use a noise-strength
$\sigma=0.2$, and include the stochastic term by generalizing ETD to
the stochastic integral, similar
to~\cite{kloeden-lord-neuenkirch-etal:2011, lord-tambue:2013}.

\subsubsection{Projection onto non-moving reference frame}

All spatially non-trivial attracting states we will be focusing on for
the deterministic Barkley model are so-called relative fixed
points---they are traveling wave solutions which move with a constant
speed along the pipe. In the reference frame moving with this velocity
they turn into fixed points, and in a periodic domain such as ours,
they are limit cycles in the lab reference frame. In order to find
these solutions with classical algorithms designed to obtain
temporally constant configurations, we project the equations
adaptively in time onto the corresponding moving reference frame, the
idea is similar to that developed in~\cite{delalamo:2009,kreilos:2014}.

In particular, in order to adaptively eliminate the object's
translation along the pipe, we project the deterministic drift of the
equation onto its part perpendicular to translation. This can be done
by realizing that $\partial_x$ is the generator of translations, so
that $n = \partial_x (q,u)$ is the direction in configuration space at
the point $(q,u)$ that points into the direction of spatial
translation. We can then project the right-hand-side of the
deterministic part of equation~(\ref{eq:model}),
\begin{equation}
  \partial_t (q,u) = b(q,u) = (b_q(q,u), b_u(q,u))
\end{equation}
with
\begin{equation}
  \begin{cases}
    b_q(q,u) = f(q,u) + D\partial_x^2q +
    (\zeta-u)\partial_xq\\ b_u(q,u) = g(q,u) + D_u\partial_x^2u
    -u\partial_xu\\
  \end{cases}
\end{equation}
onto the subspace orthogonal to $n$,
\begin{equation}
  \tilde b = b - \frac{n}{|n|^2} \langle n, b\rangle\,,
\end{equation}
where $|.|$ and $\langle .,.\rangle$ are $L^2$ norm and inner product,
so that the $\tilde b$-dynamics have no translational component. This
allows us to obtain dynamics
\begin{equation}
  \partial_t (q, u) = \tilde b(q,u)\,,
\end{equation}
that only model the deformation of objects but not their movement
speed. Note additionally that the prefactor of this projection will
yield the movement speed of the object,
\begin{equation}
  v(q,u) = \frac{\langle n, b\rangle}{|n|^2}\,,
\end{equation}
since
\begin{equation}
  \partial_t (q,u) = \tilde b(q, u) + v(q,u ) \partial_x (q,u)\,.
\end{equation}
In these projected dynamics, all states we are interested in (puff,
antipuff, decay edge, gap edge) are fixed points of the $\tilde b$
dynamics, with $\tilde b = 0$. For example, the decay edge which is a
limit cycle of $b$ is now a fixed point with $\tilde b=0$, and has a
single unstable direction corresponding to either decaying into the
laminar state, or being the minimal seed to form a puff.

Not only does this procedure allow us to treat the configurations of
interest as proper fixed points, but it also eliminates any CFL
condition from the advective term. In combination with the usage of
ETD this means that the reaction terms ($f(q,u)$ and $g(q,u)$) are the
only terms restricting the time step.

Note that we use this projection only for our deterministic
computations, as the interaction of the (spatially very rough) random
noise with the spatial derivative needed to compute the translational
component makes the projection inaccurate. For stochastic simulations,
we instead apply a spatial translation at each iteration so that that
the \emph{center of turbulent mass}, $\langle x \rangle_q = \int_0^L
x\, q(x)\,dx/\int_0^L q(x)\,dx$ remains at the domain center.

\subsubsection{Edge tracking algorithm}
\label{sec:edge-track-algor}

In order to find the \emph{stable} deterministic fixed points of the
Barkley mode, it is enough to run numerical simulations until
convergence, starting from an appropriate initial condition. For
example, in order to generate the stable puff state, we initialize
with a localized region of turbulence, which turns out to be a
configuration within the basin of attraction of the puff state for
properly chosen $r$.

For finding the \emph{unstable} fixed points, in particular the
relevant edge states between puff and laminar flow (the decay edge),
and between turbulent flow and puff or antipuff (the gap edge), we
employ edge tracking. The algorithm is implemented as follows: Define
by $B(q,u)$ the map from a configuration $(q,u)$ to its basin of
attraction $B\in\{\text{laminar}, \text{puff}, \text{turbulent},
\text{antipuff}, \text{two puffs}, \ldots\}$. Numerically, we
implement this function by integrating the deterministic dynamics
until they are stationary, and comparing their \emph{turbulent mass}
$\bar q=\int_0^L q(x)\,dx$ with that of the known fixed points. While
in general this comparison would be inconclusive (for example, a slug
might have the same turbulent mass as two puffs), it is sufficient to
identify the fixed points once the configuration is fully converged
and no longer changes.

Now, to obtain the deterministic edge state, we then integrate two
separate configurations of the system, $z_0 = (q_0, u_0)$ and $z_1 =
(q_1, u_1)$, initialized to the two fixed points between which we want
to find the edge state, for example $B(z_0)=\text{laminar}$ and
$B(z_1)=\text{puff}$. Via bisection, we iteratively approach the basin
boundary until the distance $d$ between $z_0$ and $z_1$ is below some
threshold, $d(z_0,z_1)<\Delta_\text{min}$, making sure that we also
retain that $B(z_0)=\text{laminar}$ and $B(z_1)=\text{puff}$. Since
the basin boundary is generally repulsive, $z_0$ and $z_1$ will over
time separate. Whenever they have separated too much,
$d(z_0,z_1)>\Delta_{\text{max}}$, we perform another bisection
procedure until they are again close together. This procedure is
performed until the states $z_0$ and $z_1$ converge. Effectively, the
algorithm integrates the dynamics restricted to the separating
sub-manifold, by restricting the dynamics in the unstable direction
(the separation between $z_0$ and $z_1$), while not interfering with
all other directions. The end result is a state which is stable when
restricted to the separating manifold, which corresponds to a fixed
point of the dynamics with a single unstable direction, precisely the
``saddle points'' or edge states we are interested in.

\subsubsection{Bifurcation diagram for the Barkley model} 

With the edge tracking algorithm lined out above, the schematic
bifurcation diagram shown in figure~\ref{fig:phase-diagram} can be
computed explicitly for the Barkley model by computing the relevant
fixed points and edge states for each value of $r$.

In order to efficiently compute edge states, in particular the gap
edge in the puff regime, we employed two additional techniques: First,
we used \emph{continuation} to get a good first guess for the gap edge
at a given $r$ by using the previous result for the edge computation
at a close-by $r$. Second, close to the edge we can use a
local-in-time heuristic to decide on which side of the basin boundary
a configuration is located: If its turbulent mass $\bar q$ is
increasing in time, the configuration lies towards the turbulent fixed
point, while if $\bar q$ is decreasing in time, the configuration lies
towards the puff (or antipuff). While this criterion is only true
close to the edge, it allows us to compute the unstable branch much
more efficiently.

\subsection{Asymptotic results for the Barkley model}

Here we demonstrate how the general arguments made above manifest
themselves in the deterministic Barkley model using analytical
arguments. We will focus on leading order results in $\epsilon \ll1$,
which is the parameter controlling the slow relaxation of the mean
shear $u$ in the model.

Above we have denoted front speeds by $c_{\pm}(u,r)=u-\zeta\mp
S(u,r)$, while in the notations of~\cite{barkley:2016}
$S(u,r)=\sqrt{D}s(u,r)$. Using standard
techniques~\cite{keener:2009,engel-kuehn-rijk:2021}, it can be shown
that at leading order in $\epsilon$
\begin{equation}
    s(u,r)=3\sqrt{\frac{r+u-U_0}2}-\sqrt{\frac{r+\delta}2}.
\end{equation}
One can then solve explicitly for the velocity $u_p$ at the downstream
front of a puff, though that gives a lengthy expression which we omit
here. The turbulent fixed point $(q_t,u_t)$ corresponds to the
intersection of the $u$ nullcline with the $q$ (upper branch)
nullcline defined by $q_+=1+\sqrt{\frac{r+u-U_0}{r+\delta}}$, $u_t$
being the solution to the equation $U_0-u+\kappa
(\bar{U}-u)q_{+}(u,r)=0$. The turbulent fixed point first appears at
$r_{\text{turb}}$, at the intersection of the $u$ nullcline with the
nose of the $q$ nullcline which is at $q_t=1$. This gives
\begin{equation}
    u_{t}(r_{\text{turb}})= \frac{(U_0+\kappa\bar{U})}{1+\kappa}
  \end{equation}
and
\begin{equation}
  r_{\text{turb}}= \frac{\kappa (U_0-\bar{U})}{1+\kappa}
\end{equation}
with $r_{\text{turb}}=2/3$ for our parameters.

\subsubsection{The gap edge}

We now discuss the gap edge in the limit $\epsilon\to 0$. We note that
many characteristics we describe below are identical to those of the
decay edge in this limit. We build on the analysis presented
in~\cite{barkley:2016} to make our arguments for the properties of the
gap edge.

To solve for the structure of the gap edge in the limit $\epsilon\to
0$, we may consider $u=u_t$ fixed and solve for $q$ at this fixed $u$
(this is also true for fronts of puffs and antipuffs). Then, assuming
a traveling wave solution at speed $c_g$, and moving into its
reference frame, the dynamical equation for $q$ becomes a spatial ODE:
\begin{equation}
  D\partial_{xx}q = -(c_g-u_t+\zeta)\partial_x q-f(q,u_t)
  \label{eq:one-particle}
\end{equation}
This is equivalent to a particle with position $q$, moving in a force
field $-f(q,u)$ with linear friction with coefficient $c_g-u_t+\zeta$
acting on it. The system being one dimensional, the force can be
written as a derivative of an (inverted) potential $V_r(q)$, with
$f(q,u)=\partial_q V_r(q)$, which has maxima at $q=0$ and
$q=q_+(r,u_t)$. This is an inverted potential compared to the local
dynamics for $q$, i.e $\partial_t q$ keeping $u$ fixed and considering
a spatially homogeneous solution.

\begin{figure}
  \includegraphics[height=120pt]{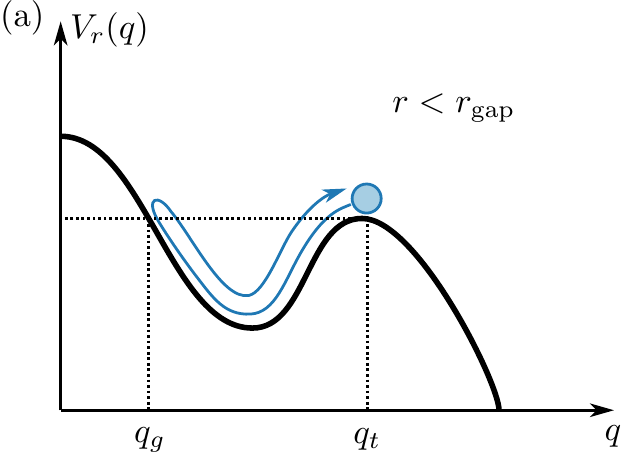}\\
  \includegraphics[height=120pt]{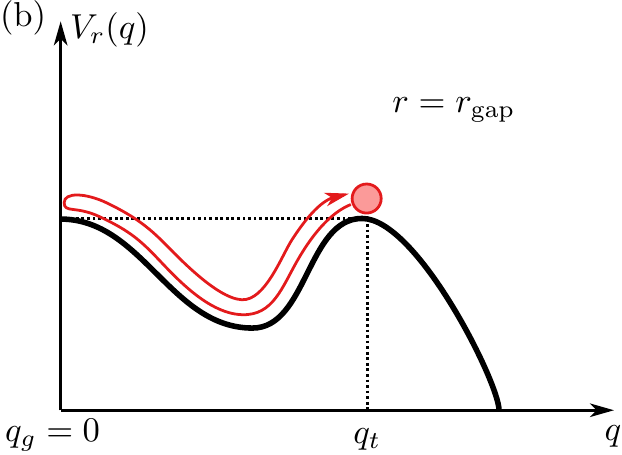}
  \caption{(a): The spatial structure of the gap edge $q(x)$ corresponds to a homoclinic trajectory of a particle moving in the potential
    $V_r(q)$, with $\partial_q V_r(q) = f(q,u)$ and $x$ playing the role of time. The particle starts at $q_t$,
    reaches $q_g$ and returns. (b) When $r=r_{\text{gap}}$, this
    homoclinic trajectory goes all the way to $q_g=0$. \label{fig:potentials}}
\end{figure}

The gap edge solution corresponds to a homoclinic trajectory of the
one particle system~\ref{eq:one-particle}: going from $q_t$ and back,
i.e $q(x\to\pm \infty)=q_t$ with zero "velocity" $\dot{q}(x\to \pm
\infty)=0$. Such a trajectory is possible as long as $q=q_t$ is the
lower maximum of the potential compared to $q=0$, which in terms of
the local dynamics of $q$ corresponds to turbulent flow being a local
minimum of the potential while laminar flow is a global minimum. From
conservation of energy in the one particle system (or time reversal
symmetry where $x$ plays the role of time), such a trajectory requires
zero friction (meaning conservative one particle dynamics), giving
$c_g=u_t-\zeta$. For $r<r_{\text{gap}}$, this situation is depicted in
figure~\ref{fig:potentials} (a). As $r$ increases, the turbulent fixed
point becomes more stable: it rises in relative height in the inverted
potential, making the homoclinic trajectory approach closer to $q=0$
as the (potential) energy of the initial condition increases, until
the laminar and turbulent maxima have identical height. At this
$r=r_{\text{gap}}$, the trajectory goes all the way to $q=0$ and the
homoclinic orbit is made of two heteroclinic orbits connecting the two
fixed points. This is the point where the gap edge and antipuff
merge, the fronts of the gap edge becoming fronts of antipuffs which
go all the way to/from $q=0$, as depicted in
figure~\ref{fig:potentials} (b). The corresponding mathematical
details are more thoroughly discussed in a general context
in~\cite{barkley:2016} Appendix A.

\subsubsection{The antipuff regime}

The transition from puffs to slugs happens when $u_t(r)=u_p(r)$, which
for our parameters gives $r_{\text{slug}}\approx 0.76$, though
$O(\epsilon)$ corrections are significant here since the range
$r_{\text{slug}}-r_{\text{turb}}$ is itself of this order. At this
$r=r_{\text{slug}}$, $S(u_t,r)=-0.13$, i.e negative as required for
the existence of an antipuff. Furthermore, solving numerically for
$u_t$ we obtain that $S(u_t(r),r\approx 0.83)=0$ so that
$r_{\text{gap}}\approx 0.83$ (again $\epsilon$ corrections are
significant here).  Note that $r=r_{\text{gap}}$ is not necessarily
the point where weak fronts of the slug~\footnote{Weak fronts of slugs are fronts followed by a refractory laminar tail, where the flow relaxes to the base laminar flow} stop existing, which instead
requires $-\zeta+S(u_t,r)>0$~\cite{barkley:2016,song:2017}.

\begin{figure}
  \begin{center}
    \includegraphics[width=\columnwidth]{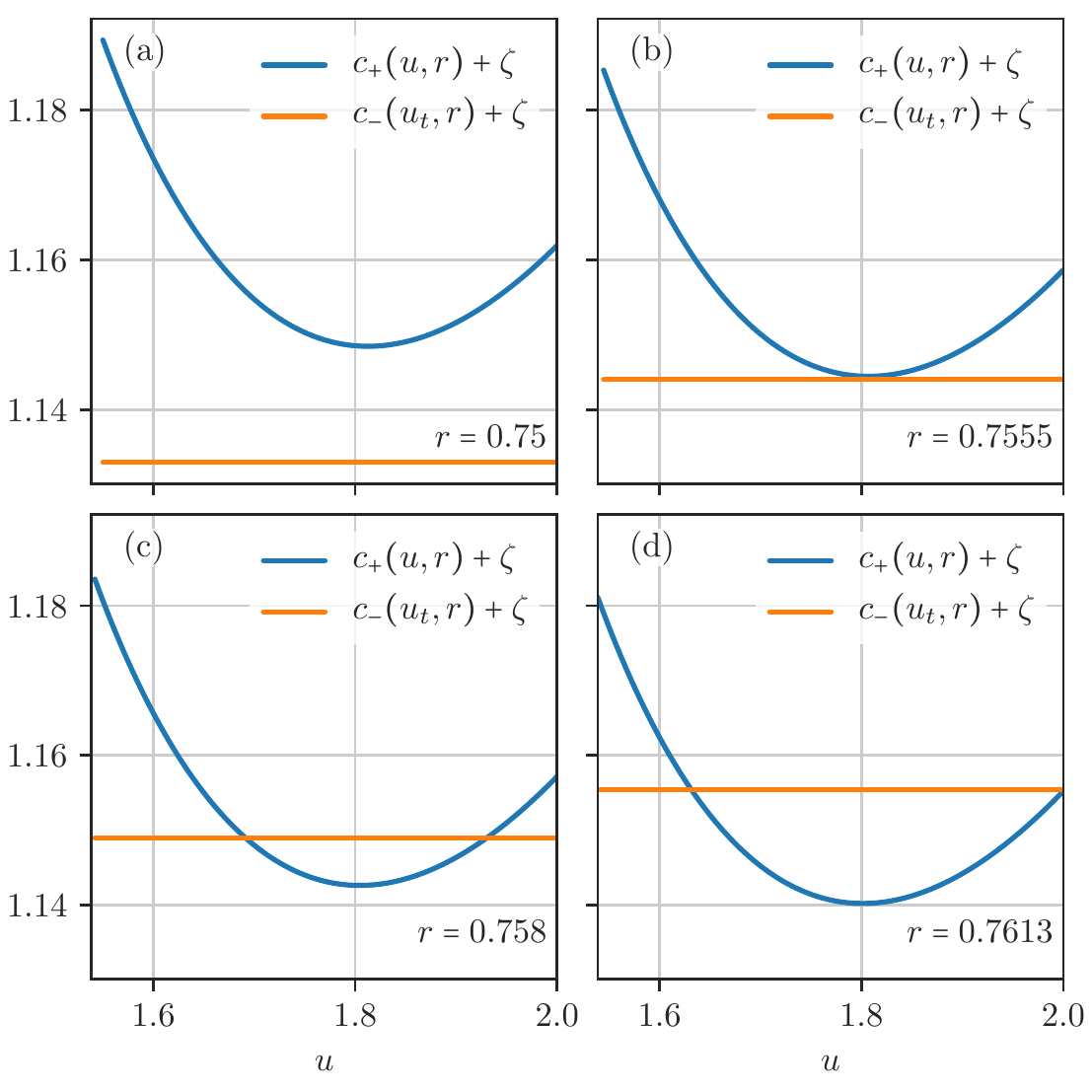}
  \end{center}
  \caption{Matching between upstream and downstream speeds for
    antipuffs in the asymptotic Barkley model. (a) Below $r=0.75$
    there is no solution for $u_{\text{ap}}$. (b) At $r\approx 0.7555$
    there is a single solution, $u_{\text{ap}}\approx 1.8$ (c) For
    $0.7555 \lesssim r\lesssim 0.7613$ there are two solutions. (d) At
    $r\approx 0.7613$ the higher velocity solution occurs at
    $u_{\text{ap}}=U_0=2$, corresponding to the identification between
    the downstream front of a puff and that of the unstable
    antipuff.\label{fig:antipuff-speed}}
\end{figure}

Although above we have focused on the case of a unique solution for
$u_{\text{ap}}$, here in the limit of $\epsilon\to 0$ there are in fact two
possible solutions. A match between front speeds of the antipuff is
first possible at $r_{\text{ap}}\approx 0.756<r_{\text{slug}}\approx
0.76$ giving $u_{\text{ap}}\approx 1.8$ (for this $r$, $u_t\approx
1.29$). In particular, at $r_{\text{ap}}\approx 0.756$ the minimum of
the curve $c_+(u,r)=u-\zeta-S(u,r)$, given by
$u=U_0-r-\frac{9D}8\approx 1.8$ touches the line $c_-(u_t,r)$, see Fig.~\ref{fig:antipuff-speed}(a,b). This
corresponds to the appearance of one stable and one unstable antipuff
in a saddle node bifurcation, as discussed below. Indeed, at higher
$0.756<r<r_{\text{slug}}$ there are two intersection points between
$c_+(u,r)=u-\zeta-S(u,r)$ and the line $c_-(u_t(r),r)$ inside the
segment $u_t(r)<u<U_0$, giving two solutions for $u_{\text{ap}}$ as in Fig.~\ref{fig:antipuff-speed}(c). At
$r=r_{\text{slug}}\approx 0.76$ the larger of the two velocities
satisfies $u_{\text{ap}}=U_0=2$ so that its downstream front is
identical to that of a puff, see Fig.~\ref{fig:antipuff-speed}(d). We will discuss how this two antipuff
scenario will manifest itself in the bifurcation diagram in the
next section. However, while it is probably realized in the Barkley
model for very small but finite $\epsilon$, its region of existence in
$r$ is minuscule, $0.756<r<0.76$, making it indistinguishable in
practice from a single antipuff appearing at $r_\text{slug}$. Thus,
we could not satisfactorily verify it in numerical simulations.

\section{Alternative scenarios for the bifurcation diagram}

Here we consider two alternative scenarios to the bifurcation diagram
presented in Fig.~\ref{fig:phase-diagram}. Remarkably, in these
scenarios there is a range of $\Re$ for which puffs and antipuffs
coexist. Both scenarios appear to be inconsistent with observations
for pipe flow, though the differences are subtle and thus could be
relevant to other wall bounded flows where puff-like and slug-like
structures occur.

The three main assumptions we have made so far are: (i) a continuous
transition from puffs to slugs, implying $\text{Re}_{\text{turb}}
<\text{Re}_{\text{slug}}$, (ii) at $\text{Re}_{\text{slug}}$
homogeneous turbulence is metastable compared to laminar flow,
corresponding to $S(u_t,\text{Re}_{\text{slug}})<0$ as can be measured
at the downstream front of a slug, and (iii) there is a unique solution
for the antipuff speed $u_{\text{ap}}$ which gives fronts of matching
speed. While the first two assumptions can be directly measured, the
third assumption is more subtle but could still be checked: it implies
that a puff continuously turns into an antipuff when viewed in the
q-u plane. That indeed appears to be the case for pipe
flow~\cite{moxey:2010}, though this issue has not be at the focus of a
dedicated study. In the following we will assume (i) is satisfied
throughout, though we are not aware of a general argument precluding a
discontinuous transition from puffs to slugs.

We begin by exploring the consequences of breaking assumption (iii)
while keeping (i) and (ii). Indeed, the equation determining the speed
of the downstream front of an antipuff does not necessarily have a
unique solution:
$u_{\text{ap}}-S(u_{\text{ap}},\text{Re})=u_t+S(u_t,\text{Re})$ can
have more than one solution, but at most two, since
$S(u_{\text{ap}},\text{Re})$ is an increasing function of
$u_{\text{ap}}$. Thus the right hand side of the equation is not
necessarily monotonic but at most has one extremum. If there are
indeed two solutions for $u_{\text{ap}}$, they correspond to the
presence of a stable and an unstable antipuff, and we will denote by
$\text{Re}_{\text{ap}}$ the Reynolds number where they first appear
together. Note that $\text{Re}_{\text{ap}}>\text{Re}_{\text{turb}}$
since an antipuff is a localized state within homogeneous turbulence.
For
$\text{\text{Re}}_{\text{ap}}<\text{Re}<\text{\text{Re}}_{\text{gap}}$,
creating a laminar pocket within homogeneous turbulence will lead to
the formation of an antipuff. Thus, the gap edge lies at the boundary
between turbulence and the stable antipuff state and the bifurcation
diagram is unchanged for
$\text{Re}_{\text{slug}}<\text{Re}<\text{Re}_{\text{gap}}$. Like
before, $\text{Re}_{\text{gap}}$ corresponds to the point where the
gap edge merges with an antipuff.

\begin{figure}
  \begin{center}
    \includegraphics[width=0.9\columnwidth]{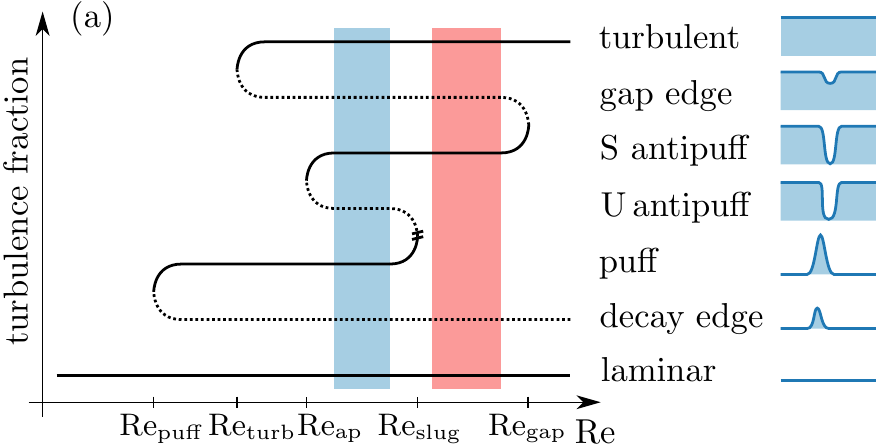}\\[2ex]
    \includegraphics[width=0.9\columnwidth]{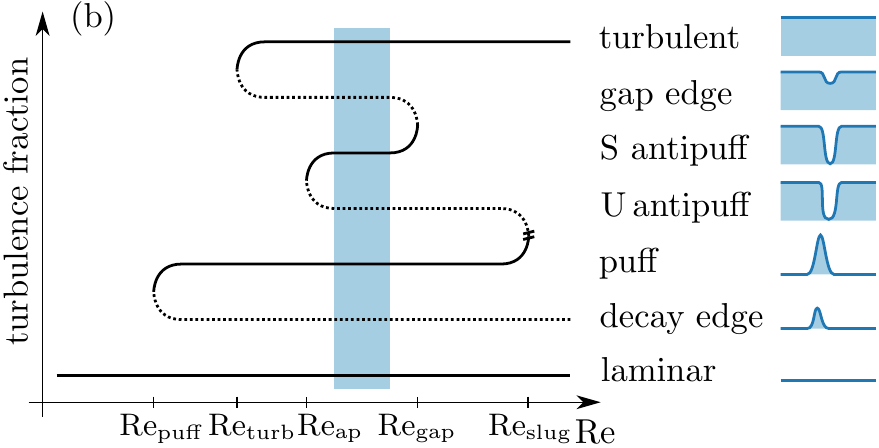}
  \end{center}
  \caption{Alternative bifurcation diagrams for transitional flow. (a): Existence of two antipuff solutions, with the unstable
    (``U antipuff'') and stable antipuff (``S antipuff'') being
    created out of a saddle-node bifurcation at
    $\Re_{\text{ap}}$. Note the co-existence of puffs and antipuffs in
    this regime. The intermittent turbulence regime is marked in red. (b) Disappearance of the gap edge before the transition to slugs: $\Re_{\text{gap}}<\Re_{\text{slug}}$. Like in (a) there are two antipuff solutions and a coexistence region between antipuffs and puffs. The intermittent turbulence regime is absent.
  } \label{fig:phase-diagram1}
\end{figure}

The stable antipuff appears at $\text{Re}_{\text{ap}}$ and disappears
at $\text{Re}_{\text{gap}}$. Thus, the unstable antipuff must
disappear at $\text{Re}_{\text{slug}}$. Indeed, at
$\text{Re}_{\text{slug}}$ $u_{\text{ap}}=U_0$ is a solution, since a
slug has matching upstream and downstream front speeds at this
$\text{Re}$. Thus, like previously, a puff turns into an antipuff at
$\text{Re}_{\text{slug}}$, but here it is the unstable antipuff.
Note that slugs, which connect the laminar base flow with homogeneous
turbulence, are still contracting (since $u_p>u_t$) for all
$\text{Re}_{\text{turb}}<\text{Re}<\text{Re}_{\text{slug}}$. However,
even though slugs are contracting, if one were to sufficiently
decrease the mean flow in the laminar region, then the laminar region
would contract to a finite length, forming the stable antipuff. The
corresponding bifurcation diagram is presented in
Fig.~\ref{fig:phase-diagram1} (a).

As a second alternative, let us briefly discuss the case where
assumption (ii) is broken while keeping assumption (i). This
corresponds to assuming $S(u_t,\text{Re}_{\text{slug}})>0$, but that
puffs still continuously turn into slugs at $\text{Re}_{\text{slug}}$. In particular, the condition
$S(u_t,\text{Re}_{\text{slug}})<\zeta$ for the existence of a weak
slug front is assumed to still be satisfied~\cite{barkley:2016}. In
this case, $\text{Re}_{\text{gap}}<\text{Re}_{\text{slug}}$ so that
stable antipuffs disappear before $\text{Re}_{\text{slug}}$. It follows that
this is also a regime with two antipuffs, breaking also assumption (iii), the unstable antipuff disappearing at
$\text{Re}_{\text{slug}}$ as before. No intermittent turbulent regime
can exist in this case. This scenario is sketched in
Fig.~\ref{fig:phase-diagram1} (b).

\section{Intermittent turbulence regime}

As stated above, we propose that the intermittent turbulence regime
corresponds to the range
$\text{Re}_{\text{slug}}<\text{Re}<\text{Re}_{\text{gap}}$, so that
laminar pockets within homogeneous turbulence observed in simulations
of pipes~\cite{moxey:2010} are in fact antipuffs which are excited
and subsequently decay. Both excitations and decays are expected to
occur through the gap edge. These laminar pockets set the fraction of
laminar flow within homogeneous turbulence, and thus have a similar
role to that of puffs for the reverse transition from turbulence to
laminar flow. Antipuffs however do not completely mirror puffs: they
can be spontaneously excited from the turbulent state, as it is not
absorbing, while on the other hand they cannot split. The fraction of
laminar flow in the homogeneous turbulent state is thus controlled by
the probabilities of antipuff excitations and decays. These vary
smoothly with $\text{Re}$, excitations becoming rarer and lifetimes
becoming shorter as the gap edge grows deeper, as indeed observed in
pipe flow~\cite{moxey:2010}, and the Barkley
model~\cite{barkley:2016}. Thus, this is not a phase transition and in
particular there is no critical point corresponding to it.

\begin{figure}
  \begin{center}
    \includegraphics[width=245pt]{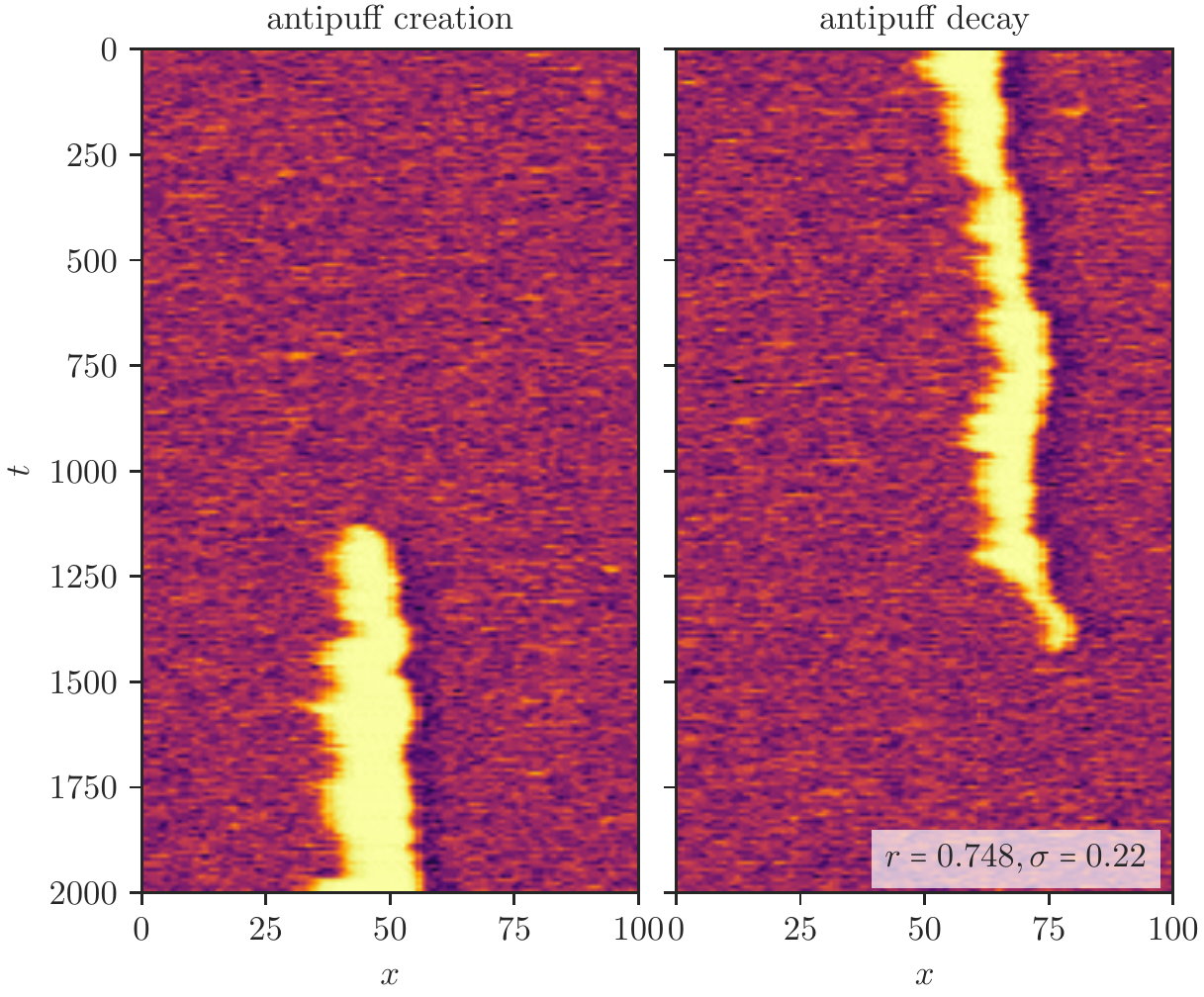}
  \end{center}
  \caption{For $r=0.748$, $\sigma=0.22$, the antipuff is long lived,
    but noise is strong enough to drive a rare transition from
    turbulent flow into the antipuff (left), and from an antipuff
    back into the fully turbulent flow (right).}
      \label{fig:antipuff-transition}
\end{figure}

We now wish to demonstrate that the laminar pockets within homogeneous
turbulence observed in the Barkley model indeed correspond to the
excitations and decays of antipuffs. We therefore consider the
stochastic Barkley model.  The stochastic model had been previously
explored for the noise level $\sigma=0.5$ in ~\cite{barkley:2016}, but
this level of noise is so high that laminar flashes are
frequent. Thus, the observation of a single creation and decay event
is hard, the pockets lifetimes are short, and multiple laminar pockets
regularly coexist. In order to isolate creation and decay of a single
stochastic laminar pocket, we perform numerical simulations at a lower
noise level, $\sigma=0.22$ and in Fig.~\ref{fig:antipuff-transition}
present a stochastic creation event (left panel), and a stochastic
decay event (right panel) both at $r=0.748$ which is lower than
$r_{\text{gap}}$ for this noise level.

\begin{figure}
  \begin{center}
    \includegraphics[width=245pt]{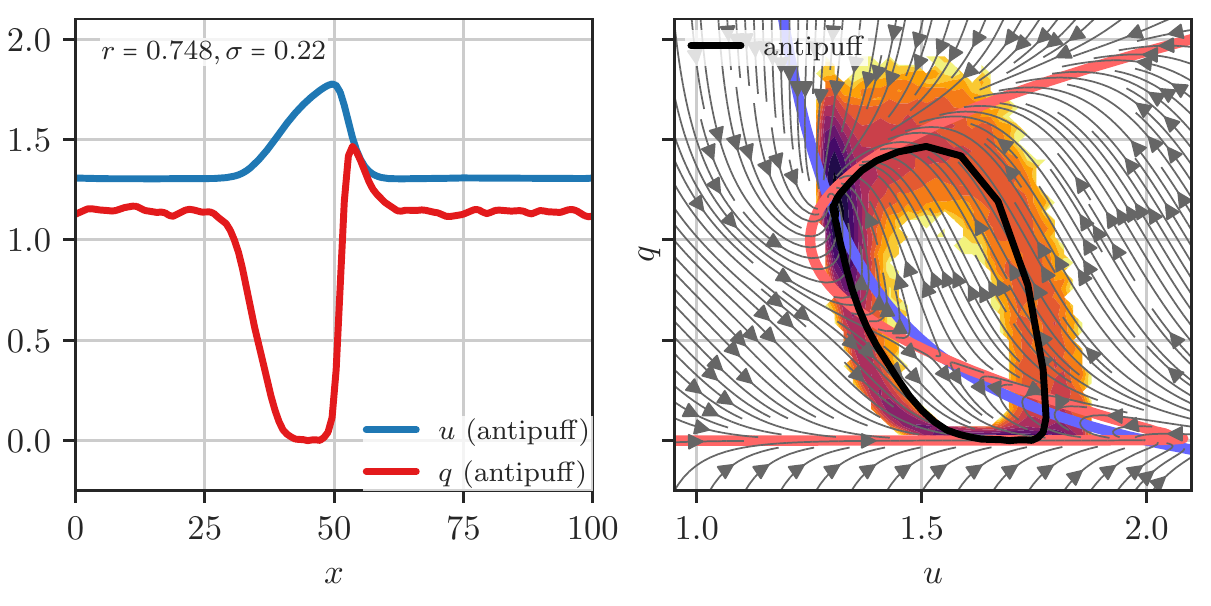}
  \end{center}
  \caption{Spatial structure (left) and $q$-$u$-plot (right) of the
    antipuff at $r=0.748$, $\sigma=0.22$. Many realizations of the
    stochastic antipuff are aligned at the downstream front to obtain
    its average form. In the $q$-$u$-plane, we superimpose a density
    plot of all considered realizations of the antipuff.}
    \label{fig:antipuff-profile}
\end{figure}

In addition, we show the profile of the stochastic laminar pocket in
Fig.~\ref{fig:antipuff-profile}, where we present both the spatial
$q$ and $u$ profile for an average pocket (right) and a $q$-$u$-plot
(left), which includes both the average as well as the density of
individual realizations. The averaging is performed by aligning the
structures in space according to the downstream front, which is
therefore sharp. Note that this smears the upstream front making the
average less representative, as is evident in the $q$-$u$-plot, since
the spatial extent of the pocket tends to vary significantly, as seen
in Fig.~\ref{fig:antipuff-transition}. The resemblance of the average
structure to the deterministic antipuff is striking. It is also
evident, for the mean as well as for individual realizations, that the
mean shear $u$ never reaches the laminar value $U_0=2$, a
characteristic feature of antipuffs.


\section{Conclusion}

We have motivated the existence of two novel states: the \emph{gap
  edge} and the \emph{antipuff} and have discussed how they fit
within a bifurcation diagram involving previously known states. Our
work motivates the study of antipuffs as well defined separate
states, as well as a search for the gap edge. It further suggests the
existence of invariant solutions which have a localized laminar region
(e.g. where streamwise vorticity is depleted) embedded in a turbulent
(vortical) flow, as those could be underlying the gap edge and the
antipuff state.

Taken together, a unified dynamical picture of the transitional regime
emerges: laminar gaps forming within homogeneous turbulence are the
mirror images of turbulent patches embedded within laminar
flow. Still, the transition from laminar flow to turbulence with
increasing $\text{Re}$ is not the mirror image of the transition from
homogeneous turbulence to laminar flow with decreasing
$\text{Re}$. This is a consequence of the absorbing nature of the
laminar base flow, which the homogeneous turbulent state does not
share. Thus, while the former transition is a proper phase transition,
the latter is not.

Finally, while we believe the bifurcation diagram we presented is
relevant for pipe flow, other alternatives are also possible. We have
presented two such alternatives here. In future work, it will be
interesting to explore their possible relevance to other wall bounded
flows and the ensuing consequences for the transition to and from
turbulence.

\textbf{\textit{Acknowledgments}} We are grateful to Dwight Barkley,
S\'ebastien Gom\'e and Laurette Tuckerman for many helpful discussions
and comments. We also thank Yariv Kafri, Dov Levine and \mbox{Grisha}
Falkovich for discussions and comments on the manuscript. TG
acknowledges the support received from the EPSRC projects EP/T011866/1
and EP/V013319/1.

\bibliographystyle{apsrev4-1}
\bibliography{bib}

\end{document}